\documentclass[osajnl,twocolumn,showpacs,superscriptaddress,11pt]{revtex4-1} 
\usepackage{amsmath,amssymb,graphicx}

\usepackage{xcolor}
\usepackage[colorlinks=true]{hyperref}

\begin{document}

\title{Low scatter and ultra-low reflectivity measured in a fused silica window}

\author{Cinthia Padilla}
\author{Fabian Maga\~{n}a-Sandoval}
\author{Erik Muniz}
\author{Joshua R. Smith}\email{Corresponding author: josmith@fullerton.edu}
\affiliation{Gravitational Wave Physics and Astronomy Center, California State University Fullerton, Fullerton, CA 92831, USA}

\author{Peter Fritschel}
\affiliation{Massachusetts Institute of Technology, Cambridge, Massachusetts 02139, USA}

\author{Liyuan Zhang}
\affiliation{California Institute of Technology LIGO Project – MS 18-34 1200 E. California Blvd. Pasadena, CA 91125}

\begin{abstract}

We investigate the reflectivity and optical scattering characteristics at 1064\,nm of an antireflection coated fused silica window of the type being used in the Advanced LIGO gravitational-wave detectors. Reflectivity is measured in the ultra-low range of 5-10\,ppm (by vendor) and 14-30\,ppm (by us). Using an angle-resolved scatterometer we measure the sample's Bidirectional Scattering Distribution Function (BSDF) and use this to estimate its transmitted and reflected scatter at roughly 20-40\,ppm and 1\,ppm, respectively, over the range of angles measured. We further inspect the sample's low backscatter using an imaging scatterometer, measuring an angle resolved BSDF below $10^{-6}$ sr$^{-1}$ for large angles (10$^\circ$--80$^\circ$ from incidence in the plane of the beam). We use the associated images to (partially) isolate scatter from different regions of the sample and find that scattering from the bulk fused silica is on par with backscatter from the antireflection coated optical surfaces. To confirm that the bulk scattering is caused by Rayleigh scattering, we perform a separate experiment, measuring the scattering intensity versus input polarization angle. We estimate that 0.9--1.3\,ppm of the backscatter can be accounted for by Rayleigh scattering of the bulk fused silica. These results indicate that modern antireflection coatings have low enough scatter to not limit the total backscattering of thick fused silica optics. 
 
\end{abstract}


\maketitle 

\section{Introduction}

Low-scatter optics are important for many scientific applications, notably ring-laser gyroscopes~\cite{Thomas1978,Chao1984}, high-power laser systems, and interferometric gravitational-wave detectors~\cite{Harry2012,Beauville2004}. Advances in ion-beam sputtering techniques to deposit dielectric multilayer coatings onto super polished substrates~\cite{Wei1989,Watkins1993,Cimma2006}, such as improved thickness control, now allow for the production of very low scatter optics and more accurate antireflection coatings. Total scatter loss of 10\,ppm or less at 1064\,nm has become standard for ion-beam-sputtered coatings~\cite{Rempe1991,Beauville2004,MaganaSandoval2012}. 

Low-scatter, low reflectivity antireflection coatings have important applications including corrective lenses~\cite{Reynolds2011}, photography, solar cells~\cite{Aiken2000,Victoria2012}, laser crystals and non-linear crystals~\cite{Stefszky2011}, and optical viewports~\cite{Massey1989}. The context of this work is interferometric gravitational-wave detection, where antireflection coatings are used to minimize reflections and scatter from the non-reflective secondary surfaces of the interferometer optics and from both surfaces of viewports used to transmit laser beams into and out of the vacuum system.

A worldwide network of second-generation gravitational-wave detectors, including Advanced LIGO~\cite{Harry2010}, Advanced VIRGO~\cite{AdvVirgo}, KAGRA~\cite{Somiya2012}, and GEO-HF~\cite{Willke2006} is currently under construction. These interferometers require exquisite displacement sensitivity, of order 1$\times$10$^{-20}$m/$\sqrt{\mathrm{Hz}}$ around 100\,Hz, in order to directly measure the weak effects that gravitational waves from astrophysical systems have on test masses on earth. 

Scattered light can decrease the sensitivity of gravitational-wave detectors in several ways. Each detector's primary optics are made of fused silica substrates with ion-beam-sputtered dielectric coatings~\cite{Harry2012} to produce highly reflective, anti-reflective, or beam-splitting optical surfaces. Optical loss from light scattered by the highly reflective primary optics used in optical cavities can reduce the laser power build-up in the interferometers and decrease their shot-noise-limited sensitivity. The use of non-classical light, such as squeezed light, to improve the quantum-noise limited sensitivity of the interferometers can be severely degraded by light scattering losses from the optics used to prepare and inject the squeezed states~\cite{Aasi2013a,Abadie2011a,MaganaSandoval2012}. Finally, scattered light from the primary and auxiliary optics, including viewport windows that are used to pass beams into and out of the vacuum system, can couple back into the interferometer adding non-linear noise~\cite{Ottoway2012}. 

This paper presents a characterization of the light scattering properties of an ion-beam-sputtered anti-reflective coated viewport. This optic is found to have very low scatter and therefore is suitable for use in Advanced LIGO.

\section{Sample and preparation}

\begin{figure}
\centerline{\includegraphics[width=.7\columnwidth]{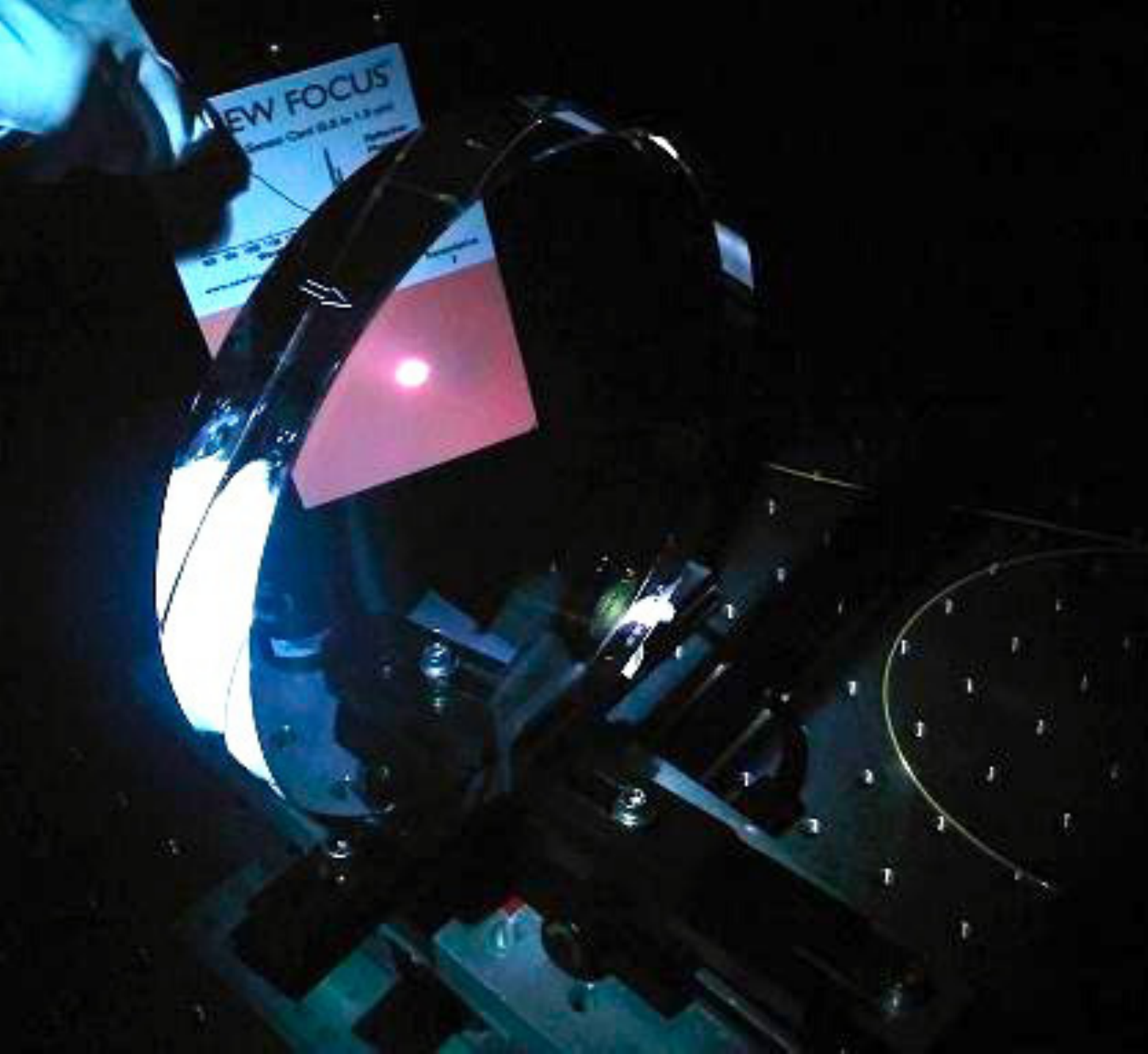}}
\caption{Fused silica viewport ESW03 shown in roomlight mounted on the rotation stage of the Fullerton Imaging Scatterometer. A laser beam is passing through both surfaces, but is not visible in the photo. The arrow indicates the forward surface. Both surfaces have identical antireflective coatings.}\label{fig:viewport-photo}
\end{figure}

A room light image of the sample investigated here, Research Electro-Optics, serial number ESW03, is shown in Figure~\ref{fig:viewport-photo}. The substrate is made of high-quality fused silica (Corning 7980, 0A) with a six inch diameter and a thickness of one inch. Both flat optical surfaces are super polished with 10/5 scratch-dig surface quality and have ion-beam-sputtered antireflection coatings that were specified to provide very low reflectivity for 1064\,nm (goal of 10\,ppm). 


Prior to measurement, the sample surfaces were cleaned of dust and other contaminants by drag-wiping with optic tissues and ultra-pure methanol. To further reduce contamination from dust, the scatter measurements were conducted in a cleanroom environment.    

\section{Measurements}

The sample was measured to have ultra-low reflectivity at 1064\,nm for small angles of incidence. Reflectivity measurements performed by the vendor, Research Electro Optics, using a 5.7\,W Nd:YAG laser and calibrated photodetector, yielded values of 8\,ppm and 6\,ppm for the front (arrowed), and back surfaces, respectively.  Later measurements at Caltech, using a 1\,mm beam at 1064\,nm and an angle of incidence $<1^{\circ}$ found reflectivities of 30\,ppm for the front side and 14\,ppm for the back side. It is not clear why these two measurements differed, but it may have to do with nano-layers of contaminants that have been observed on other optics~\cite{harald}.


Two types of Angle Resolved Scatter (ARS) measurements were performed on the sample. Transmitted and reflected scatter were measured with a photodiode-based commercial scatterometer and followup measurements of the low backscatter were made with a CCD-camera based imaging scatterometer.  

For both experiments, the laser source was 1064\,nm and the scatter was quantified according to the standard Bidirectional Scattering Distribution Function~\cite{Stover2012},
\begin{equation}\label{eqn:bsdf}
BSDF\left(\theta_s\right) = \frac{P_s}{P_i\Omega\cos{\theta_s}},
\end{equation}
where $P_i$ is the laser power incident on the sample and $P_s$ is the scattered light power collected by a detector subtending solid angle $\Omega$ at polar scattering angle $\theta_s$ in the plane of the laser beam. The BSDF is also referred to as BRDF and BTDF in the following to explicitly denote scattered light measured in reflection and transmission, respectively. 

From these BSDF measurements, the Total Integrated Scatter (TIS) of the sample for the wavelength and incidence angle used were estimated by integrating the cosine-corrected BSDF assuming independence of scatter on the azimuthal angle, following the steps described in reference~\cite{MaganaSandoval2012}. However, it should be noted that although this technique is widely used, it is not strictly correct, since a smooth sample illuminated normally by linearly polarized light will not exhibit constant scatter for all azimuthal angles at a given polar angle~\cite{Stover2012}. 


\subsection{Photodiode ARS measurements}

\begin{figure}
\centerline{\includegraphics[width=\columnwidth]{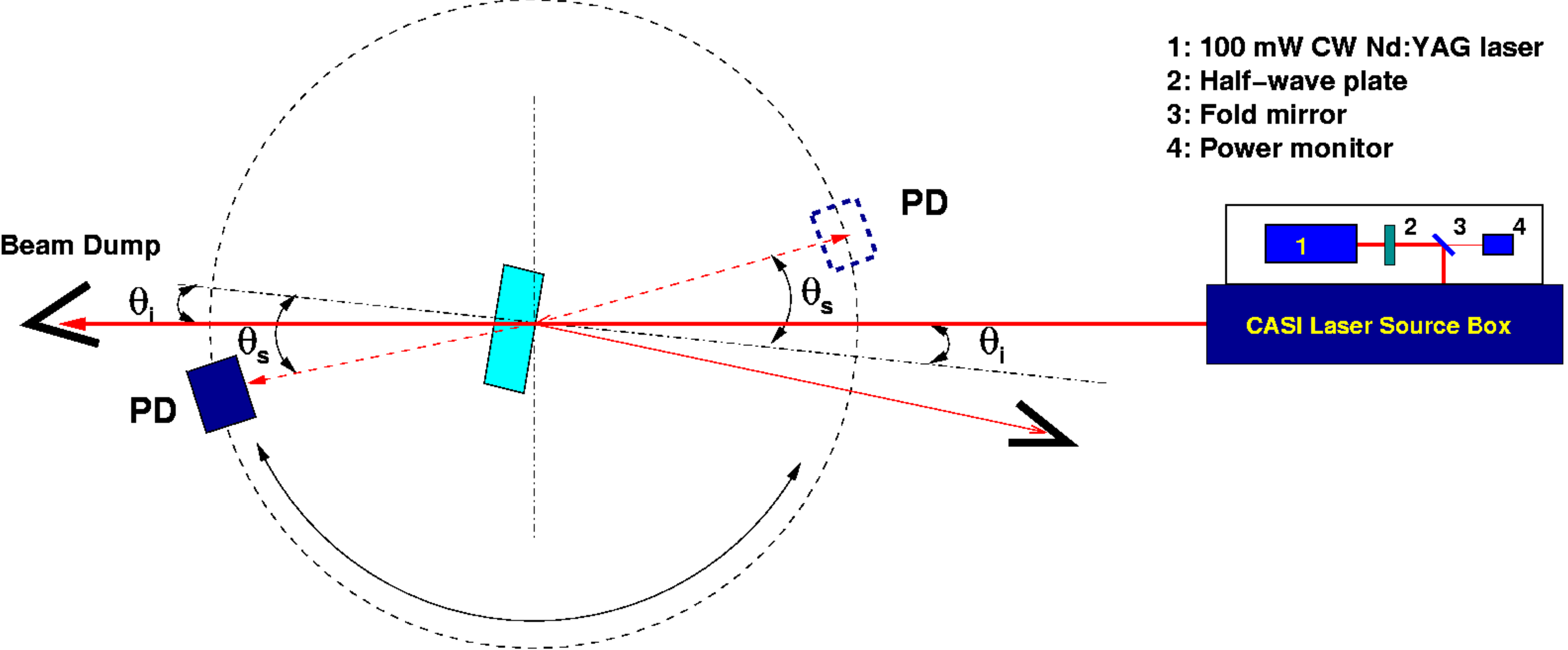}}
\caption{Setup for the Caltech Angle Resolved Scatter measurements.}\label{fig:casi}
\end{figure}


\begin{figure}
\centerline{\includegraphics[width=\columnwidth]{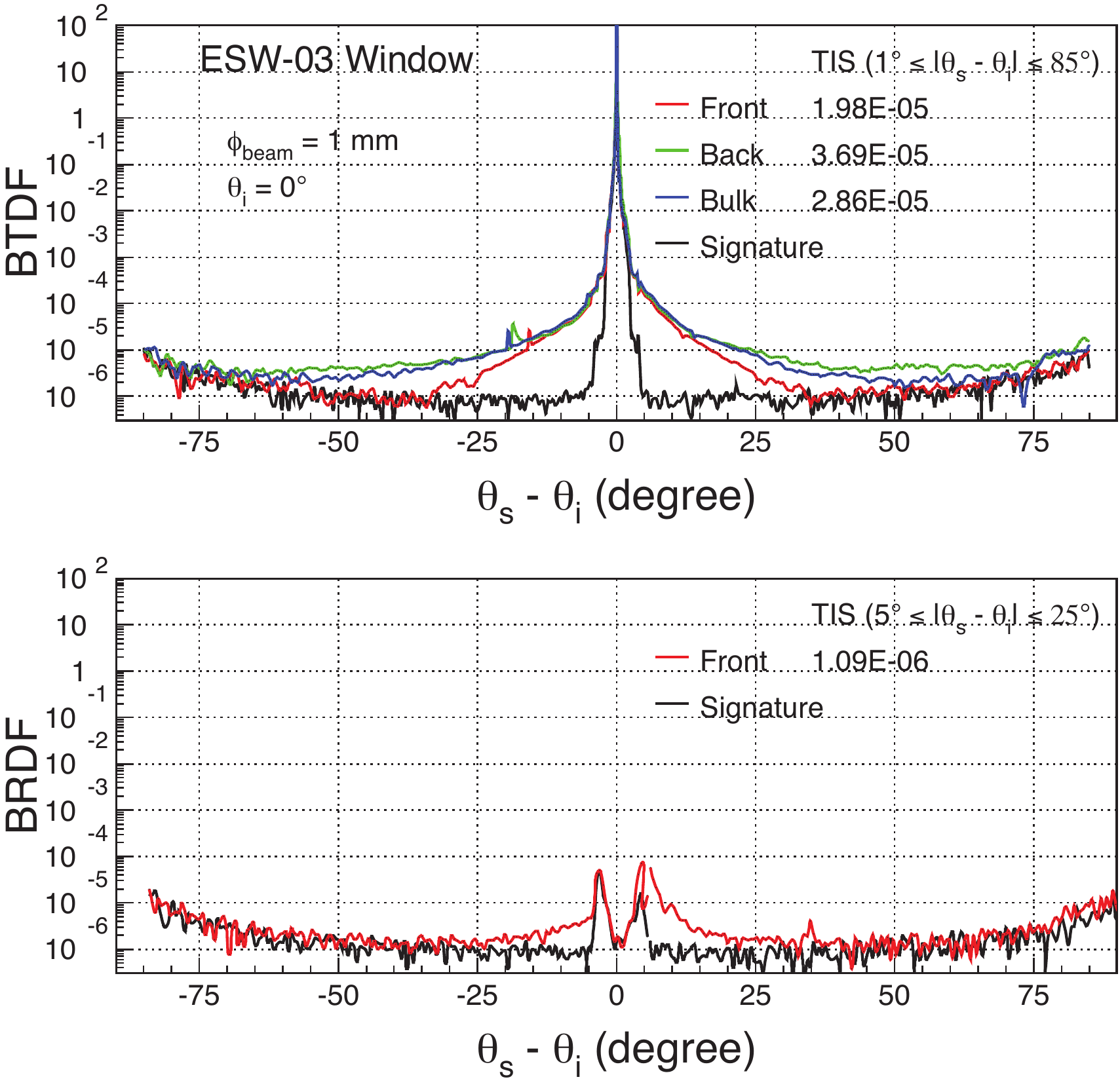}}
\caption{Scatter measurements of the ESW-03 viewport from the transmitted (BTDF, above) and reflected (BRDF, below) sides, made using the modified CASI scatterometer at Caltech. Total integrated scatter estimates are also indicated in the legend. The calibration precision for these BRDF values is estimated at better than 5\%.}\label{fig:casi-brdf}
\end{figure}

Angle resolved scatter measurements for the viewport sample were carried out at Caltech using a modified version of the commercial Complete Angle Scatter Instrument (CASI), manufactured by Schmitt Measurement Systems, Inc., and shown in Figure~\ref{fig:casi}. The CASI system originally had a He-Ne laser installed in its source box. To test scatter at 1064\,nm, this source was replaced with a Nd:YAG laser (CrystaLaser CL1064-100, 100 mW CW), along with a half-wave plate to allow changes of polarization. The beam diameter at the sample is about 1\,mm. This system is capable of measuring scattered light over all polar angles, $-90^{\circ}<\theta_s<90^{\circ}$ from normal to the sample surface, in the plane of incidence, by rotating a photodetector around the sample. A map of scatter over the sample surface can also be obtained by fixing the angle of the detector while scanning the beam position on the sample. To avoid biasing the results for samples with low scatter, both transmitted and reflected beams have to be carefully trapped with beam dumps. 

Figure~\ref{fig:casi-brdf} shows the ARS results for a representative area on the ESW03 window measured on the transmission (BTDF) and reflection (BRDF) sides. For the BTDF measurement, three scans were taken by aligning the front surface (facing laser source), middle of the window (bulk) and back surface at the rotation axis respectively. However, additional tests revealed that the scatterometer is not able to distinguish scatter from the front, back, and bulk surfaces at small angles (where the three curves match closely). 

Also shown in Figure~\ref{fig:casi-brdf} is the calculated TIS by integrating cosine corrected and background (signature) subtracted BTDF and BRDF within $1^{\circ}<|\theta_s|<85^{\circ}$ and $5^{\circ}<|\theta_s|<25^{\circ}$ for transmission and reflection sides respectively. The two sides, $\theta_s>\theta_i$ and $\theta_s<\theta_i$ are averaged in the calculation. These results indicate that most of the scatter (20-40\,ppm taking into account the overlapped BTDF at small angles) is in transmission, while only about 1\,ppm is backscatter.  

\subsection{Imaging ARS measurements}


\begin{figure}
\centerline{\includegraphics[width=\columnwidth]{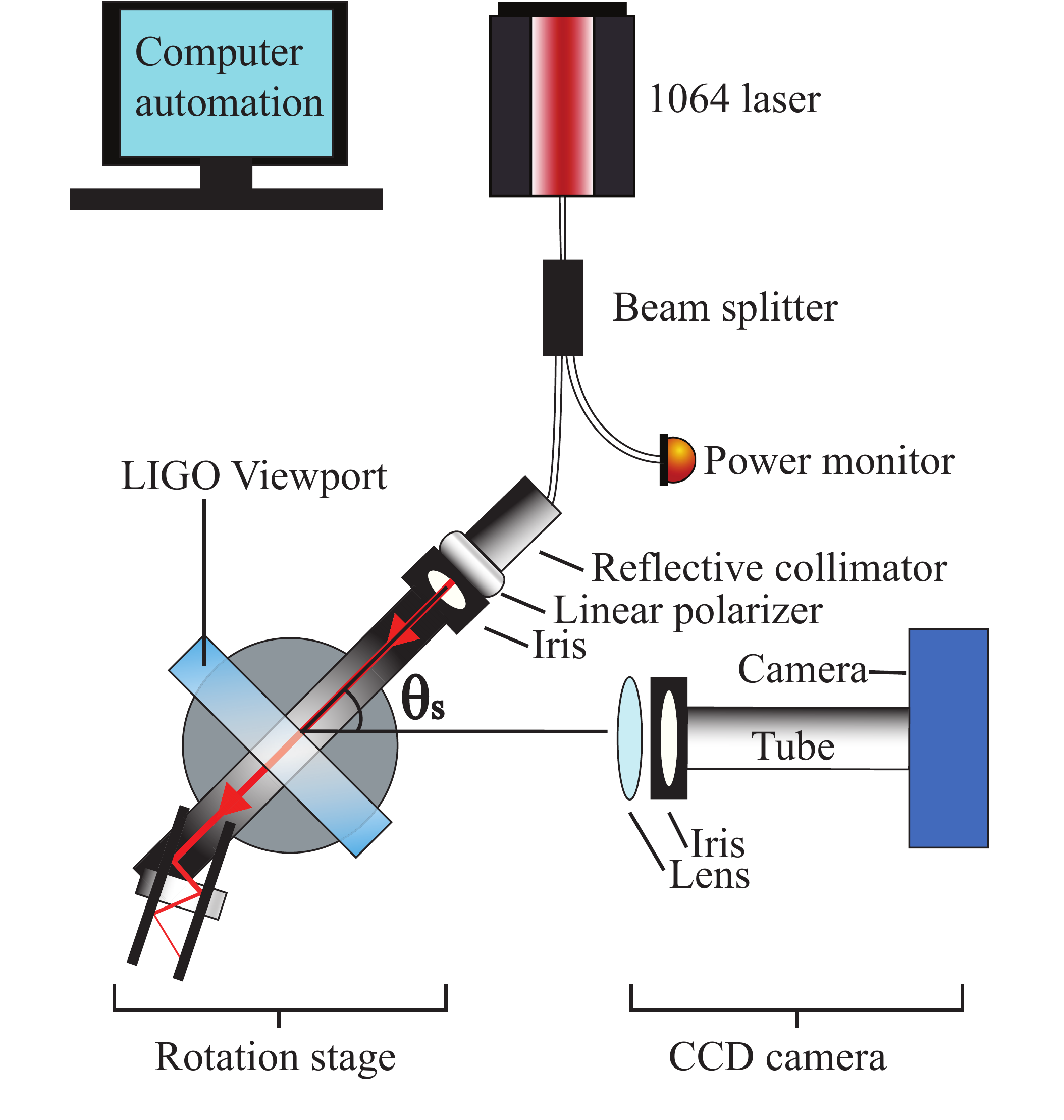}}
\caption{Setup of the Fullerton imaging scatterometer that is described further in the text and in ~\cite{MaganaSandoval2012}.}\label{fig:csuf-scatter-setup}
\end{figure}

The Fullerton Imaging Scatterometer (FIS) is shown in Figure~\ref{fig:csuf-scatter-setup} and described in detail in reference~\cite{MaganaSandoval2012}. Here the setup is briefly recounted and and differences from the previous setup are highlighted. The light source is a 1064\,nm linearly polarized Nd:YAG laser (Innolight Mephisto INN401). This is coupled into a 90:10 fiber beamplitter. The fiber's low-power output is connected to a power monitor and its high-power output is connected to a reflective collimator that collimates the beam to 8\,mm radius. This light passes through a thin-film linear polarizer set to pass horizontal polarization. The beam is then spatially truncated by an iris with an opening of 4\,mm, 
and is incident on the sample at near normal incidence. Because the viewport is antireflection coated, the reflected beams were extremely weak and were directed back to the iris to dump their power. Care was taken to dump the transmitted beam power using a multi-reflection trap made of black welder's glass.  The fiber launch, collimator, polarizer, iris, sample, and beam dumps are all mounted on a motorized rotation stage so that when the sample is rotated a fixed angle of incidence is maintained and all beams remain dumped. A single converging lens and iris are used to form an image of a 1.3" x 1.3'' region of the sample on the 1024 x 1024 pixel CCD camera (Apogee Alta U6 low-noise cooled astronomical camera). Much care is taken with optical bandpass filters, tubes, and beam blocks to minimize other sources of light that could reduce the sensitivity of the CCD images to scatter from the sample. 


\begin{figure}
\centerline{\includegraphics[width=\columnwidth]{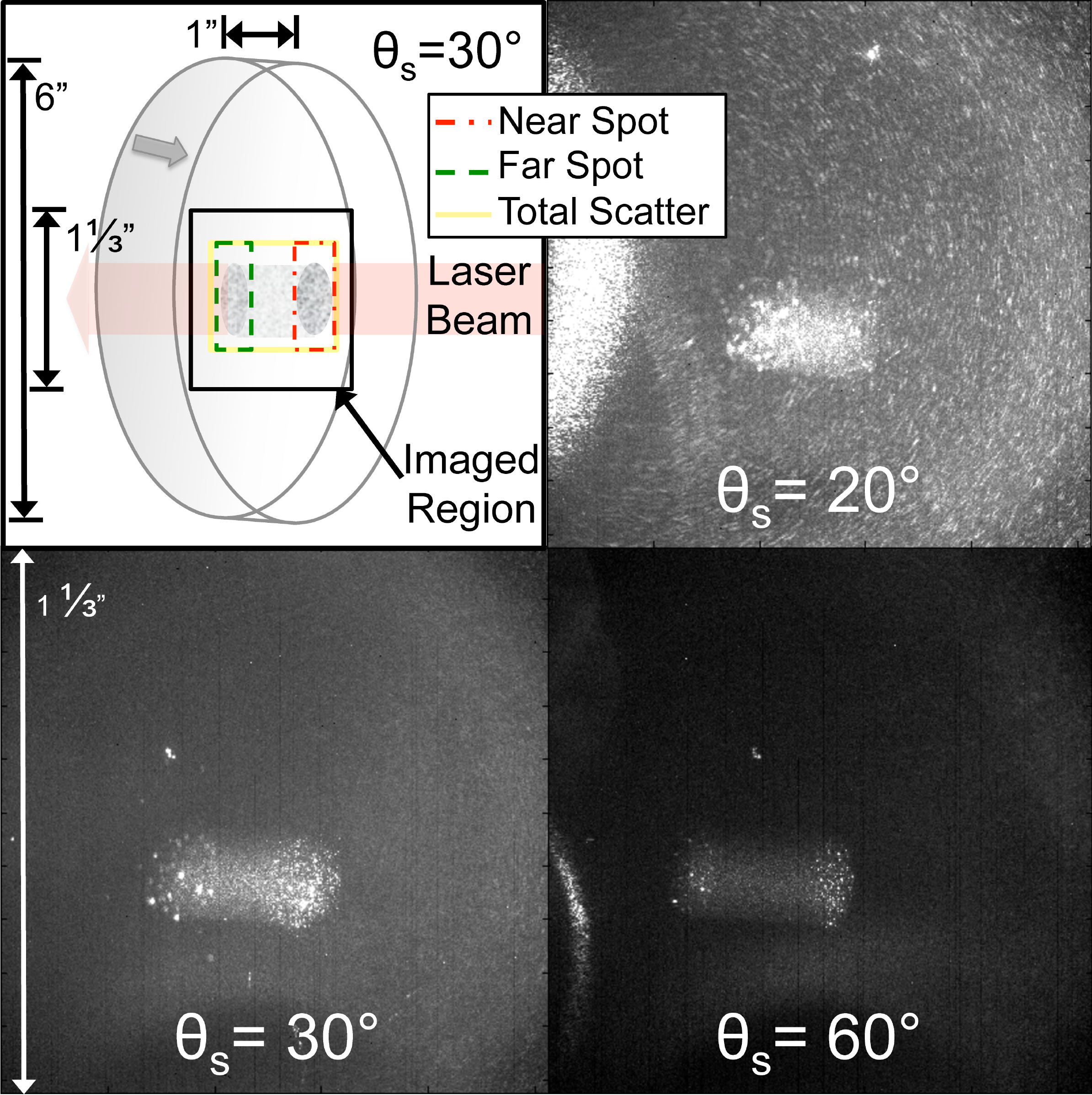}}
\caption{Upper left: Cartoon of the viewport from the perspective of the CCD camera for a scattering angle of 30 degrees. The CCD images a square 1.3" region that contains the front surface scattering, the back surface scattering, and the bulk scattering. Other panels: CCD images for three separate scattering angles, 20, 30, and 60 degrees, showing the scattering spots on the front and back surfaces, and the bulk scattering from the illuminated volume.}\label{fig:scatter-images}
\end{figure}

\begin{figure}
\centerline{\includegraphics[width=\columnwidth]{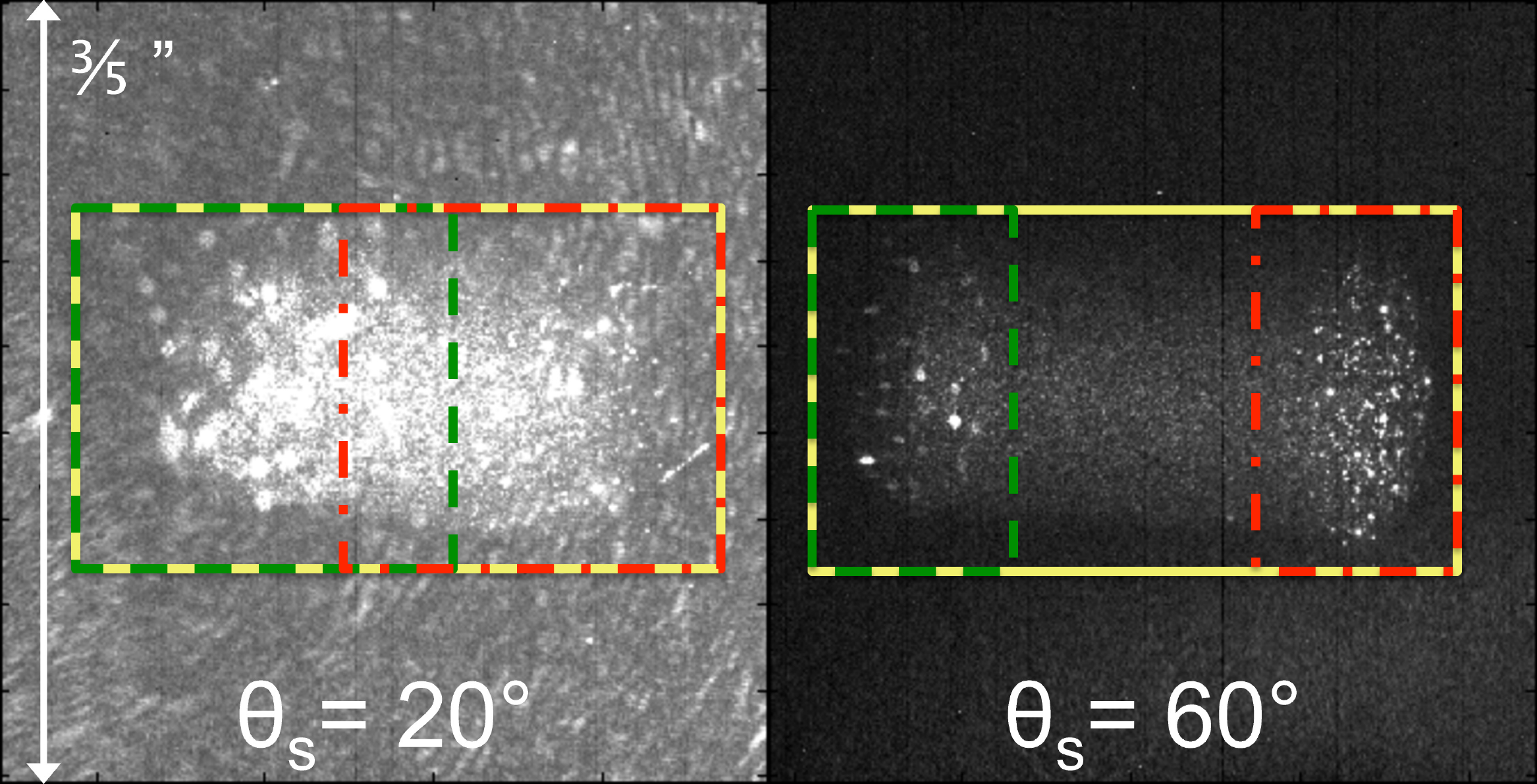}}
\caption{Zoomed-in images showing the regions of interest used to capture the total scattered light (R1) and isolate scatter from the near surface (R2) and far surface (R3). For small viewing angles, such as $\theta_s=20^{\circ}$, shown on the left, the near and far surface scatter is highly overlapped with the bulk scatter. For wider angles, such as $\theta_s=60^{\circ}$, shown on the right, the near and far surface scatter is spatially separated.}\label{fig:scatter-regions}
\end{figure}

The scatter measurements were supervised by a LabView automation Virtual Instrument. This moves the rotation stage to the desired scattering angle, measures the laser light at the power monitor, and exposes the CCD chip for imaging. Using this procedure, images and incident power measurements were collected at fixed scattering angles $0^{\circ}\leq\theta_s\leq90^{\circ}$, in one degree increments of $\theta_s$. Exposure times were set to be as long as possible such that no pixels saturated in the region of interest (see Figure~\ref{fig:scatter-regions}), typically between 30 and 50 seconds. After the 90 images are collected, the same procedure is followed, but with the laser off. These ``dark images'' are subtracted from the scattering images to reduce noise and hot pixels. The subtracted images are then analyzed using a custom Matlab script that calibrates the images (see below) and calculates the scattered light power in the regions of interest by summing all of the pixel values.  

Figure~\ref{fig:scatter-images} shows a cartoon of the viewport sample from the perspective of the CCD camera and indicates the imaged region. As shown, the bulk scattering and back surface scattering are imaged through the front surface. Also shown are full 1024 x 1024 pixel CCD images for three different scattering angles, each using the same black/white scaling. In these images, scattering from the front and back surfaces is visible as a constellation of points and bulk scattering from the illuminated volume is seen as a uniform ``glow" with roughly the same brightness. At small angles, of $20^{\circ}$ or less, the images are bright, the front and back scattering spots spatially overlap, and the bulk scattering is difficult to distinguish from the surface scattering. For angles above $30^{\circ}$ the front and back scattering spots are spatially separated and the bulk scattering can be clearly seen, appearing roughly as bright as the surface scattering. 

Figure~\ref{fig:scatter-regions} shows a zoom of the CCD images, and the regions of interest used to estimate the scatter from the front, back, and bulk scattering. Region R1 encompasses all scatter from the sample bulk and surface within the radius of the laser beam. Regions R2 and R3 are centered on the front and back spot, respectively, with their starting position set by hand. Their centers follow the beam spot motion as the sample rotates, and decrease in width with the cosine of the scattering angle. Although R2 and R3 are centered on the front and back surface spots, they capture a significant amount of bulk scatter at all angles, and overlap spatially for scattering angles below 30 degrees. Still, a lower limit estimate of the bulk scatter can be obtained by subtracting the measured scattered power in R2 and R3 from the total scattered power in R1 and converting this to BSDF. Research is ongoing to implement elliptical regions of interest in future analyses. 




The calibration technique used to convert CCD counts into physical units (BRDF) is the same as that described in~\cite{MaganaSandoval2012}. A diffuse scattering target (Spectralon Diffusion Material, 100 × 0.01200 disk, SM-00875- 200) is illuminated with the 1064\,nm laser at normal incidence. The resulting diffuse scatter is measured by a calibrated power meter and the BRDF is calculated following Equation~\ref{eqn:bsdf} using the measured incident power, the scattered light power at several scattering angles, and the solid angle subtended the power meter, giving a value close to $1/\pi$\,sr$^{-1}$ for $0^{\circ}\leq\theta_s\leq90^{\circ}$. Then images are taken of the same diffusing target at the same scattering angles using the CCD camera (and an additional $T=1/256$ neutral density filter to reduce the light power). Relating the BRDF measurements to the images gives a calibration factor. For these measurements, a value of $F=3.20\times 10^{-14}$ W sec Counts$^{-1}$ sr$^{-1}$ was used.


The major problem with this method is that it relies on the linearity of the CCD camera, the shutter timing, and dark noise subtraction in extrapolating from measured light intensities proportional to $BRDF_{\mathrm{Spectralon}}/T_{\mathrm{filter}}=1/(256\pi)\approx 10^{-3}$ from the Spectralon sample (with filter) to much smaller intensities proportional to $BRDF=10^{-7}$ from the viewport sample. Measuring linearity over this range is challenging because of limitations to calibrated power meters. Taking the known factors into account gives a calibration error of up to 50\%.  

The results and references presented below indicate that the bulk scattering from fused silica produces a BRDF that is of the same order of magnitude as the BRDF expected from the optical surfaces of low scattering samples. Research is now underway to calibrate the FIS CCD camera by directly measuring the Rayleigh scattering in fused silica and comparing it with theoretical and measured values available in the literature. The first steps toward this are presented in Appendix~\ref{appendix}.


\begin{figure}
\centerline{\includegraphics[width=\columnwidth]{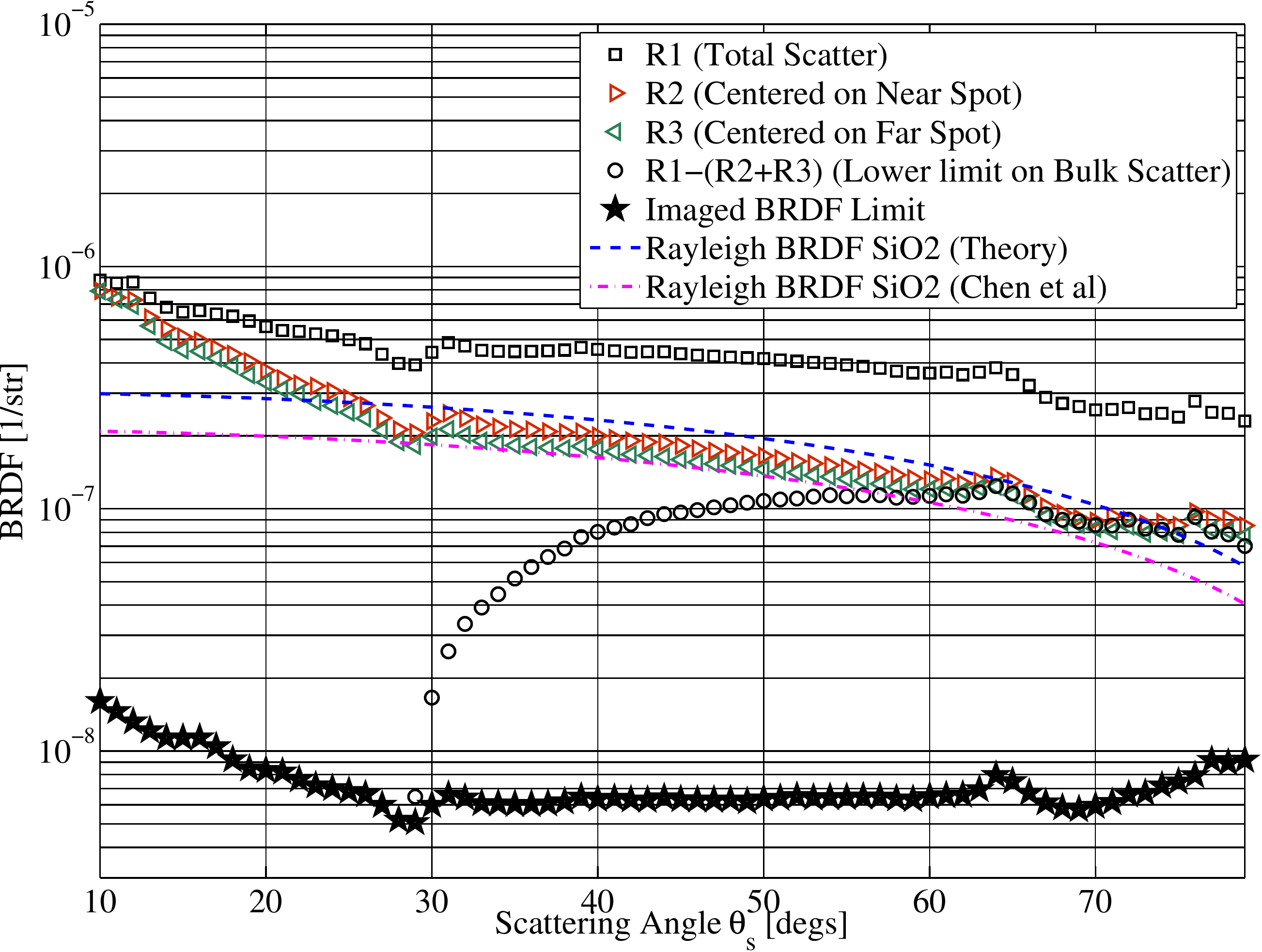}}
\caption{BRDF versus scattering angle for the viewport sample. Total scatter from region R1 (squares), which includes the near surface, far surface, and bulk scatter, is in the range 2--9$\times10^{-6}$sr$^{-1}$. Regions R2 (right-pointing triangle) and R3 (left pointing triangle), are centered on scatter from the near spot and far spot, respectively, and a lower limit on the BRDF from bulk scatter is estimated by subtracting R2 and R3 from R1 (circles). Dotted lines indicate the BSDF expected from Rayleigh scattering according to~\cite{Chen2011}. The instrument signature BRDF (stars) is a few ten times lower than the total scatter. The calibration precision for these BRDF values is better than 50\%.}\label{fig:bsdf}
\end{figure}

Figure~\ref{fig:bsdf} shows the measured BRDF for the viewport sample. The total backscatter (from region R1) is very low, 2--9$\times10^{-6}$sr$^{-1}$ over the angles measured, but an order of magnitude above the instrument signature. This BRDF is comparable to high-quality highly-reflecting ion beam deposited coatings on superpolished substrates~\cite{MaganaSandoval2012,Watkins1993}. The regions R2 and R3, centered on the near and far surface scattering, respectively, have nearly equal BRDF, but these values are not a clean indication of the scatter from the separate surfaces because the associated regions spatially overlap for small angles, and contain a significant amount of bulk scattering for all angles. 

Also shown in Figure~\ref{fig:bsdf} is a lower limit on bulk scattering, estimated by subtracting the BRDF from the near and far surface scattering from the total BRDF. This has a value of roughly $10^{-7}$sr$^{-1}$ for angles above 40$^{\circ}$, and for larger angles, approaches the expected BSDF for Rayleigh scattering (dotted lines) converted from the intensity ratio values in~\cite{Chen2011}. The bulk scattering lower limit does not match the calculated Rayleigh scattering BRDF for smaller angles because there the near and far surface regions spatially overlap in the images. At large angles, the lower limit is close to the BRDF expected from Rayleigh scattering even though, as seen in Figure~\ref{fig:scatter-regions}, R1 and R2 contain nearly as much bulk scatter as region in between. If this were corrected for, the measured data would thus be about a factor of two above the dashed lines. This discrepancy could be due to one of several small angles present in the setup. The input light polarization was only accurate to a few degrees since it was aligned by hand using the polarization axis markings on a half-inch polarizer. Also, as can be seen from Figure~\ref{fig:scatter-regions}, the incident beam was not entirely normal to the sample, instead it had a vertical angle of incidence of $2.2^{\circ}$. To ensure that the bulk scatter was caused by Rayleigh scattering, additional measurements were made to test the scattering intensity dependence on input polarization. These are presented in Appendix~\ref{appendix}. 

\begin{table}
  \caption{Integrated scatter estimates for regions of the analyzed images (total, front spot, back spot, and subtracted regions to give lower limit on Rayleigh scattering). These values are compared to TIS for Rayleigh scattering in fused silica calculated from theoretical and measured scattering coefficients from Reference~\cite{Chen2011} multiplied by the viewport sample thickness ($TIS=\alpha_{sc}t$).}\label{tab:tis}
  \begin{center}
    \begin{tabular}{cccc}
    \hline
    Region & $\theta_s$ Range & $\Omega$ (sr) & TIS (ppm)\\
    \hline
Total (R1) & $10-80\deg$ & 1.62$\pi$ & 1.26  \\
R2 (front spot) & $10-80\deg$ & 1.62$\pi$ & 0.62  \\
R3 (back spot) & $10-80\deg$ & 1.62$\pi$ & 0.56 \\
R1$-$(R2$+$R3) & $10-80\deg$ & 1.62$\pi$ & 0.21 \\
\hline
Rayleigh~\cite{Chen2011} (meas.) & $0-180\deg$ & $4\pi$ & 1.8\\
Rayleigh~\cite{Chen2011} (theo.) & $0-180\deg$ & $4\pi$ & 2.5\\
    \hline
    \end{tabular}
  \end{center}
\end{table}

Table~\ref{tab:tis} shows integrated scatter estimates for the scattering regions described above, with an assumed independence of scatter on the azimuthal angle, calculated as described in~\cite{MaganaSandoval2012}. Also shown are the TIS estimated by Chen \emph{et. al.} based on their measured and theoretical values~\cite{Chen2011}. A significant fraction of the total backscatter can be attributed to Rayleigh scattering in the fused silica bulk material.      
	
\section{Discussion}

The viewport measured here exhibits low forward scatter, very low backscatter, and ultra low reflectivity. For this sample, Rayleigh scattering from the fused silica substrate is on par with scatter from the two optical surfaces. This means that ion-sputtered antireflective coatings have nearly reached the limit where scattering from the bulk will dominate backscattering from the surface and coatings, for thick optics (viewports, compensation plates, etc). The scattered measured here was restricted to angles greater than one or a few degrees, however very small angle scattering from viewports is also of interest for gravitational-wave detector optics, and should be measured.

\acknowledgements{This work is supported by National Science Foundation Awards PHY-0970147 and PHY-1255650 and by the Research Corporation for Science Advancement Cottrell College Science Award \#19839. E. Muniz is supported by (STEM)$^2$, funded by the US Department of Education (Grant\# P031C110116-12). The authors thank K. Wanser and G. Childers (CSU Fullerton) for useful discussions about Rayleigh scattering. We thank our colleagues in the LIGO Scientific Collaboration for fruitful discussions about this research and for review of this manuscript.}

\appendix


\newpage
\section{Confirming Rayleigh scattering as cause of imaged bulk scattering}\label{appendix}

\begin{figure}[h]
\centerline{\includegraphics[width=\columnwidth]{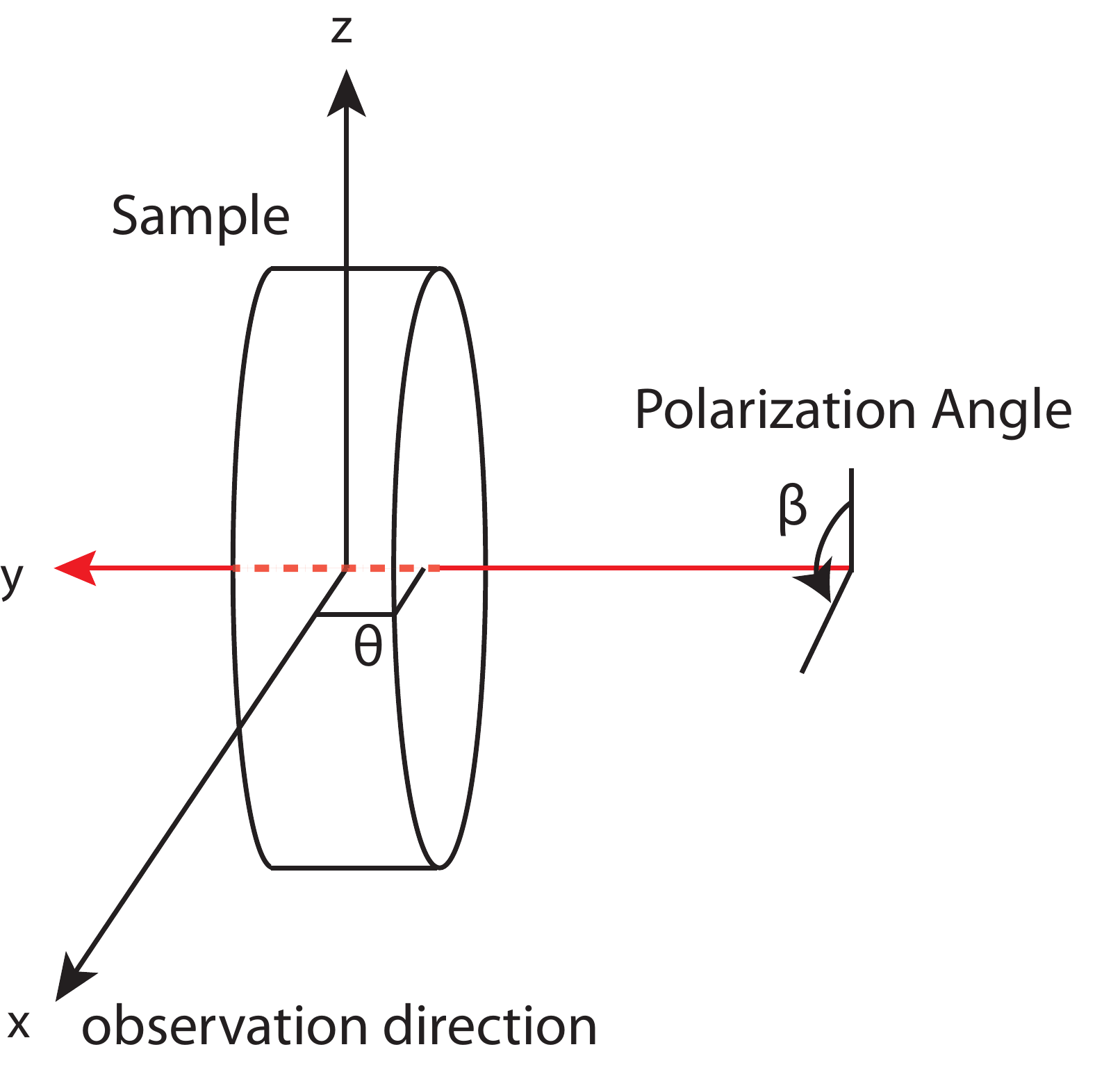}}
\caption{Diagram of the setup used to check for the functional dependence of Rayleigh scattering intensity viewed through the side of the viewport versus input polarization angle $\beta$.}\label{fig:Rayleigh_Diagram}
\end{figure}

\begin{figure}
\centerline{\includegraphics[width=\columnwidth]{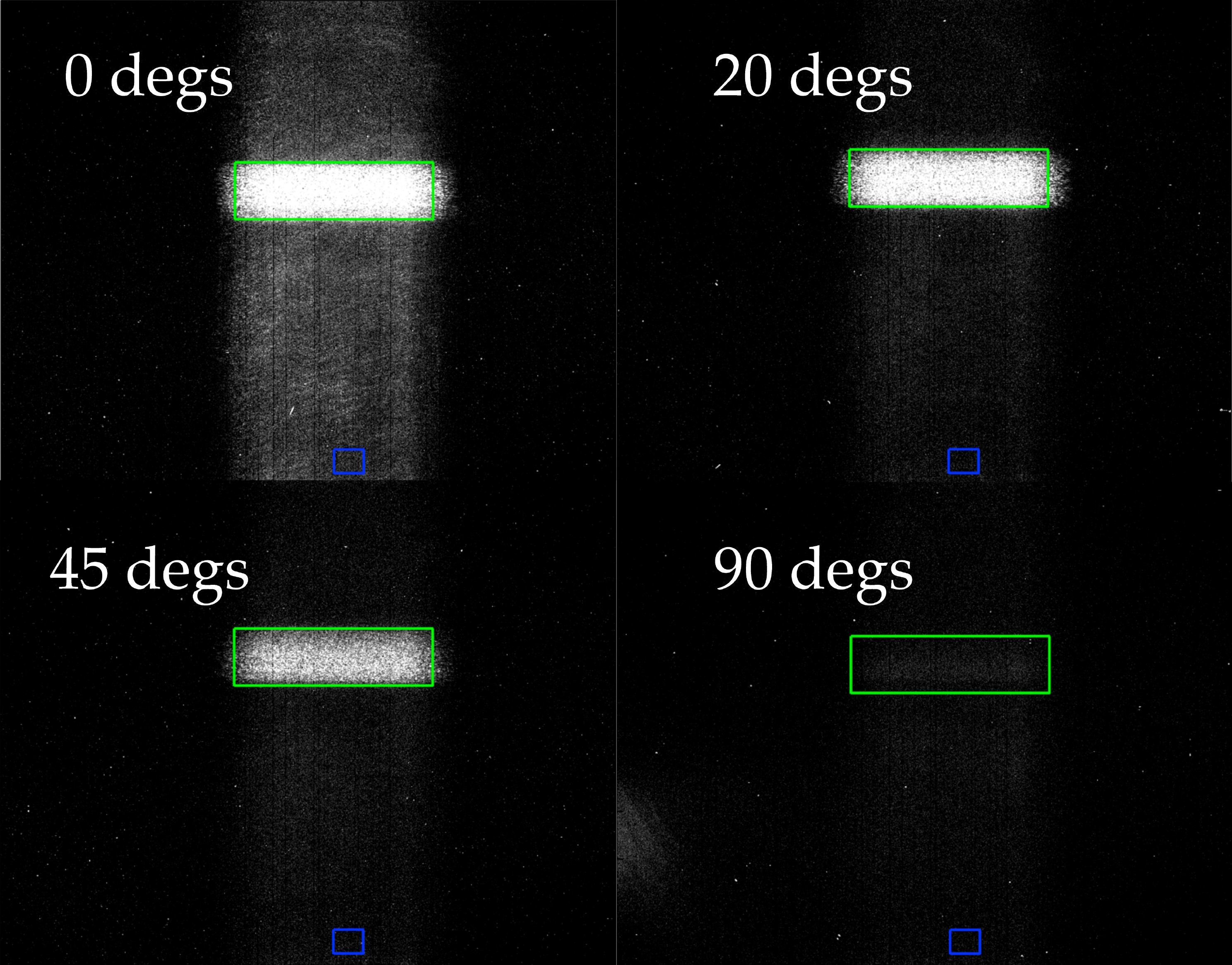}}
\caption{Imaged scattered light intensity through the side of the viewport for vertically, 20 degrees, 45 degrees, and horizontally polarized input light. The scatter RoI is the wide rectangle around the scattering and the background RoI is the small box at the bottom of each image.}\label{fig:Rayleigh_Images}
\end{figure}

An additional experiment was performed to confirm that the bulk scattering from the viewport measured by the Fullerton Imaging Scatterometer was dominated by Rayleigh scattering of the fused silica material. A diagram of the experimental setup, which follows that used in~\cite{Chen2011}, is shown in Figure~\ref{fig:Rayleigh_Diagram}. The input beam is normally incident on the viewport and the transmitted beam is dumped. The CCD camera observes the bulk of the optic through its barrel, perpendicular to the input beam and parallel to the table. Beam blocks are used to block the light from the beveled edges of the optic barrel, and only the central 1cm of the optic is imaged. The linear polarization angle of the input beam is adjusted by rotating a linear polarizer on a graded rotation mount, and power is kept high by rotating the input light polarization to match the axis of the polarizer. Images of the bulk scattering for three different input polarization angles are shown in Figure~\ref{fig:Rayleigh_Images}.

\begin{figure}
\centerline{\includegraphics[width=\columnwidth]{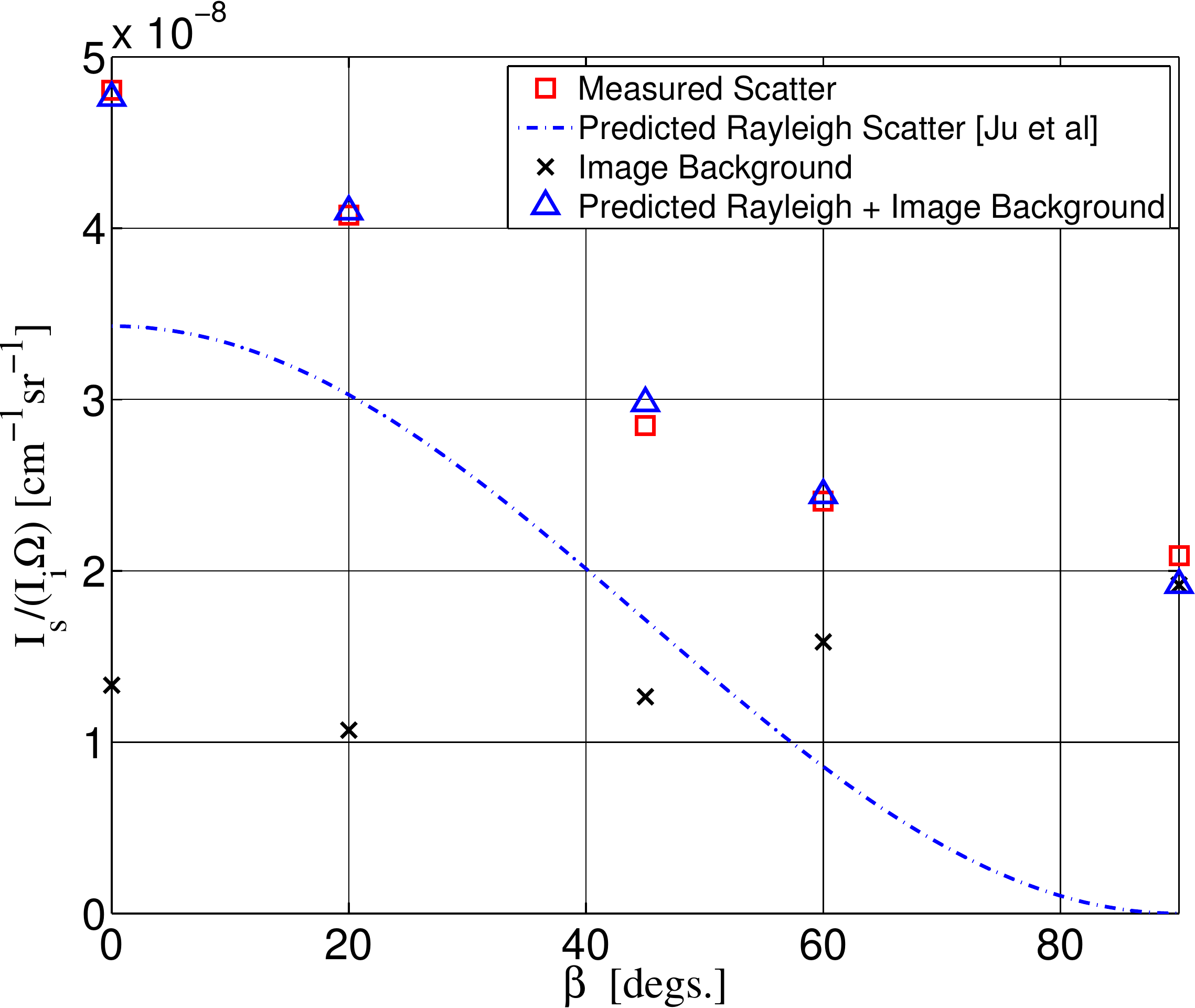}}
\caption{Squares represent the scattering intensity ratio measured through the side of the viewport versus varying input beam polarization angle $\beta$. The dashed line is the expected intensity function for Rayleigh scattering based on measurements in Ju et. al.~\cite{Chen2011} and the $\cos(\beta)^2$ dependency. For each image, an estimate of the background noise, produced by taking the intensity ratio in a 50x50 pixels region (vertically aligned with the scattering, but far below the beam in the image) and scaling it by the ratio of pixels contained in the scattering RoI to those contained in the background RoI, is marked with an x. The measured scatter matches well the sum of the expected Rayleigh scattering and the image background noise. }\label{fig:Rayleigh_Intensity}
\end{figure}  

The measured intensity ratio, based on the CCD calibration above, is shown in Figure~\ref{fig:Rayleigh_Intensity}. Also shown is a measurement of the background noise in each image. This background is likely due to additional stray light in the setup that is not present in the dark images, and does not spatially overlap with the beam, as would be expected for e.g., from additional Rayleigh scattering due to depolarization in fused silica~\cite{Brinkmeyer1983}. Also shown is the expected intensity ratio for Rayleigh scattering based on the $I\propto\cos(\beta)^2$ functional dependence and the measured maximum scattering (at $\beta = 0$) from~\cite{Chen2011}. The measured scatter matches well the sum of the expected Rayleigh scattering ratio and image background noise, confirming that it is Rayleigh scattering.

\end{document}